\documentclass[11pt,twoside]{article}

\usepackage{asp2004}
\usepackage{epsf}
\usepackage{psfig}
\usepackage{lscape}

\begin{document}
\title{Massive Stars and their Compact Remnants in High-mass X-ray Binaries}
\author{L. Kaper \& A. Van der Meer}
\affil{Astronomical Institute, University of Amsterdam, Kruislaan 403,
 1098 SJ Amsterdam, The Netherlands}

\begin{abstract} 
In a high-mass X-ray binary (HMXB) a massive star interacts with a
neutron-star or black-hole companion in various ways. The
gravitational interaction enables the measurement of fundamental
parameters such as the mass of both binary components, providing
important constraints on the evolutionary history of the system, the
equation of state of matter at supra-nuclear density, and the
supernova mechanism. The stellar wind of the massive star is
intercepted by the strong gravitational field of the compact
companion, giving rise to the production of X-rays. The X-rays
increase the degree of ionization in a small to very extended region
of the surrounding stellar wind, depending on the X-ray
luminosity. This has observable consequences for the structure and
dynamics of the accretion flow. In this paper we concentrate on the
fundamental paramaters of the most massive HMXBs, i.e. those with an
OB supergiant companion, including some systems exhibiting
relativistic jets (``microquasars'').
\end{abstract}

\section{High-mass X-ray binaries}

In a high-mass X-ray binary (HMXB) a massive OB-type star is in close
orbit with a compact X-ray source, a neutron star or a black hole. The
X-ray source is powered by accretion of material originating from the
OB star, transported by the OB-star wind or by Roche-lobe
overflow. The majority ($\ga 80$~\%) of the HMXBs are Be/X-ray
binaries with relatively wide ($P_{\rm orb}$ weeks to several years)
and eccentric orbits. Most Be/X-ray binaries are transients, the X-ray
flux being high when the compact star in its eccentric orbit passes
through the dense equatorial disk around the Be star (Van den Heuvel
\& Rappaport 1987, Negueruela, these proceedings). About a dozen HMXBs
host a massive OB-supergiant companion (about 10 to over
40~M$_{\odot}$), in a relatively tight orbit ($P_{\rm orb}$ several
days) with an X-ray pulsar or black-hole companion (e.g. Kaper
2001). Some of these X-ray binaries include a dense accretion disk and
produce relativistic jets (e.g. Fender 2005).

HMXBs mark an important though short phase (on the order of 10,000
year) in the evolution of the most massive binaries. The compact
companion is the remnant of the initially most massive star in the
system that exploded as a supernova (or as a gamma-ray burst). Due to
a phase of mass transfer, the secondary became the most massive star
in the system before the primary supernova, so that the system
remained bound (Van den Heuvel \& Heise 1972). A consequence, however,
is that HMXBs are runaways due to the kick velocity exerted by the
supernova (Blaauw 1961, Kaper et al. 1997, Van den Heuvel et
al. 2000). When the secondary starts to become a supergiant, a second
phase of mass transfer is initiated, first through an enhanced stellar
wind, later by Roche-lobe overflow, which results in the production of
X-rays by the compact companion. With increasing mass transfer rate,
the system will enter a phase of common-envelope evolution causing the
compact object to spiral into the OB companion. In the relatively wide
Be/X-ray binaries the spiral-in likely results in the removal of the
envelope of the Be companion, and after the supernova a bound (or
disrupted) double neutron star remains, like the Hulse-Taylor binary
pulsar PSR~1913+16 (or a neutron star -- white dwarf system, if the mass
of the Be companion is less than $\sim 8$~M$_{\odot}$). In close
HMXBs (orbital period less than about a year) the compact object will
enter the core of the OB companion which will become a so-called
Thorne-Zytkow object (Thorne \& Zytkow 1977), a red supergiant with a
high mass-loss rate.  These objects have been predicted on
evolutionary grounds, but have so far not been recognized as such (for
an extensive review on binary evolution, see Van den Heuvel 1994).

In case the system hosts an X-ray pulsar, its orbit can be very
accurately determined through pulse-timing analysis. When also the
radial-velocity curve of the OB supergiant is obtained, the mass of
the neutron star and that of the massive star can be derived with
precision, given an estimate of the system inclination (cf.\ Rappaport
\& Joss 1983). The neutron star mass carries information on the
formation mechanism (i.e. the supernova), as well as on the equation
of state (EOS) of matter at supra-nuclear density. Evolutionary
calculations by Timmes et al. (1996) show that in single hydrogen-rich
stars with a mass less than 19~M$_{\odot}$ the collapsing Fe core has
a mass of about 1.3~M$_{\odot}$, while stars with a higher initial
mass produce an Fe core of about 1.7~M$_{\odot}$. Therefore, one would
predict a bimodal mass distribution of the neutron stars in these
systems. 

The EOS, i.e. the relation between pressure and density in the
neutron-star interior, can so far only be studied on the basis of
theoretical models. It is not yet possible to produce the required
extremely high density (about an order of magnitude higher than an
atomic nucleus) in accelerator experiments. Theoretical predictions of
the EOS come in two flavors: the so-called ``soft'' and ``hard''
equations of state. The hardness of the EOS depends on the fraction of
bosons formed in the neutron-star interior, which unlike the fermions
(neutrons) do not contribute to the Fermi pressure that helps to
sustain gravity. The harder the EOS, the higher the mass a neutron
star can have. If a neutron star has a mass higher than the maximum
mass allowed by a given EOS, this EOS is ruled out as only one EOS can
be the right one. Obviously, this relatively simple measurement has
major implications for our understanding of the properties of matter
at supra-nuclear density.

The properties of the OB star in a HMXB give insight into the
evolutionary history of the binary system: its current mass provides a
constraint on the initial mass of the primary, its surface may show
evidence of nuclearly processed material coming from the primary, and
its rotation rate has been altered by the angular momentum content of
the gained mass. Also, the space velocity of the system carries
information on the amount of mass lost during the supernova explosion
(Nelemans et al. 1999, Ankay et al. 2001).

Contrary to the Be/X-ray binaries, a few OB-supergiant systems host a
black-hole companion (e.g. Cyg~X-1, Gies \& Bolton 1982), which
suggests that black holes are formed by the most massive stars
only. The majority of the about twenty known black-hole candidates
have a low-mass companion (soft X-ray transient or X-ray nova,
e.g. McClintock \& Remillard 2005), but only the three black-hole
candidates known in HMXBs are persistent X-ray sources. Some of the
black holes in X-ray novae appear to be linked to hypernovae believed
to power gamma-ray bursts, based on the detection of r-processed
elements at the surface of the low-mass companions (e.g. Israelian et
al. 1999).

\section{OB supergiant systems}

Table~1 lists the basic properties of the HMXBs with OB supergiant
companions in the Milky Way and the Magellanic Clouds. Most sources
contain an X-ray pulsar: the pulse period is short and the X-ray
luminosity high (near the Eddington limit $L_{X} \sim
10^{38}$~erg~s$^{-1}$) in systems undergoing Roche-lobe overflow due
to the higher accretion rate. The latter systems also have circular
orbits, while the wind-fed systems have eccentricities up to $e =
0.45$ (GX301-2) and an X-ray luminosity $L_{X} \sim
10^{35}-10^{36}$~erg~s$^{-1}$.

The X-ray source drastically increases the degree of ionization of the
surrounding stellar wind. Hatchett \& McCray (1977) predicted that the
ionizing power of the X-ray source would cause the orbital modulation
of ultraviolet resonance lines formed in the stellar wind. The
Hatchett-McCray effect has been detected in several systems (e.g. Van
Loon et al. 2001), and recently in the wind-fed system 4U1700-37 for
which the original prediction was made (see Iping et al., these
proceedings). Van der Meer et al. (2005a) find evidence for the
ionization zone in 4U1700-37 through X-ray spectroscopy carried out
with XMM-{\it Newton}. A secondary effect of the presence of an X-ray
ionization zone is the development of a strong shock (called a
photo-ionization wake) at its trailing border, where fast wind
collides with the stagnant flow inside the ionization zone (Blondin et
al. 1990, Kaper et al. 1994). In Roche-lobe overflow systems the high
X-ray luminosity makes that only in the X-ray shadow behind the OB
supergiant a normal stellar wind can develop, a so-called shadow wind
(e.g. Blondin 1994, Kaper et al. 2005).

\begin{table}[ht]
\caption[High-mass X-ray binaries with OB supergiant
companion]{High-mass X-ray binaries with OB supergiant companion in
the Milky Way and the Magellanic Clouds (ordered according to right
ascension). The name corresponds to the X-ray source, the spectral
type to the OB supergiant. For the systems hosting an X-ray pulsar the
masses of both binary components can be measured (given an estimate of
the inclination of the system). The last five systems most likely
contain a black-hole candidate; for the galactic sources relativistic
jets have been detected. The system parameters were taken from Reig et
al. 1996 (2S0114+650); Van der Meer et al. 2005b (SMC~X-1, LMC~X-4,
Cen~X-3); Barziv et al. 2001 (Vela~X-1); Kaper, Van der Meer \&
Najarro 2005 (GX301-2); Van Kerkwijk et al. 1995 (4U1538-52); Clark et
al. 2002 (4U1700-37); Cox, Kaper \& Mokiem 2005 (4U1907+09); Cowley et
al. 1983 (LMC~X-3); Hutchings et al. 1987 (LMC~X-1); McSwain et
al. 2004 (LS5039); Hillwig et al. 2004 (SS433); Herrero et al. 1995
(Cyg~X-1). The rapid X-ray pulsars are found in Roche-lobe overflow
systems. Notes: $^a$ A spin period of 2.7~h is proposed by Corbet et
al. (1999).}
\smallskip
\begin{tabular*}{\textwidth}{@{\extracolsep{\fill}}llcccc}
\hline
\smallskip
Name & Sp.\ Type & M$_{\rm OB}$ & M$_{\rm X}$ & P$_{\rm orb}$ & P$_{\rm pulse}$ \cr
     &           & (M$_{\odot}$) & (M$_{\odot}$) & (d) & (s) \cr
\hline
2S0114+650 & B1 Ia         & $\sim$16      & $\sim$1.7     & 11.6  & 860$^a$ \cr
SMC X-1    & B0 Ib         & 15.5          & 1.1           &  3.89 & 0.71 \cr
LMC X-4    & O7 III-IV     & 15.6          & 1.3           &  1.40 & 13.5 \cr
Vela~X-1   & B0.5~Ib       & 23.9          & 1.9           &  8.96 & 283  \cr
Cen~X-3    & O6.5 II-III   & 19.7          & 1.2           &  2.09 & 4.84 \cr
GX301-2    & B1.5 Ia$^{+}$ & $>$40         & $>$1.3        & 41.5  & 696  \cr
4U1538-52  & B0 Iab        & 16.4          & 1.1           &  3.73 & 529  \cr
4U1700-37  & O6.5 Iaf+     & $\sim$58      & $\sim$2.4     &  3.41 &      \cr
4U1907+09  & early B I     & $\sim$27      & $\sim$1.4     &  8.38 & 438  \cr \hline
LMC X-3    & B3 Ve         & $\sim$6       & 6--9          &  1.70 &      \cr
LMC X-1    & O7-9 III      & $\sim$20      & 4--10         &  4.22 &      \cr
LS5039     & O6.5 V((f))   & 20--35        & 1.4           &  4.43 &      \cr
SS433      & A3-7 I        & 10.9          & 2.9           & 13.08 &      \cr
Cyg X-1    & O9.7 Iab      & 17.8          & 10            &  5.60 &      \cr
\hline
\end{tabular*}
\end{table}

The mass of the OB supergiant primary and the compact companion in a
HMXB can be accurately measured when the system hosts an X-ray
pulsar. Knowledge of the orbital inclination is essential; in systems
showing an X-ray eclipse the inclination must be larger than $i \sim
65^{\circ}$. For Roche-lobe overflow systems a valid assumption is
that the OB supergiant is in corotation with the orbit, which provides
a strong constraint on the inclination. In eclipsing systems the
radius of the OB supergiant can be derived from the duration of the
X-ray eclipse. The mass ratio sets the size of the Roche lobe (e.g.\
Eggleton 1983); it turns out that the measured radii of the OB
supergiants are in very good agreement with the estimated size of the
Roche lobe (Kaper 2001).

Earlier studies (e.g.\ Conti 1978, Rappaport \& Joss 1983) suggested that
the OB supergiants in HMXBs are too luminous for their masses. E.g.\ the
O6.5 giant companion of Cen~X-3 has a mass of 20~M$_{\odot}$, while its
luminosity corresponds to that of a star of more than 50~M$_{\odot}$.
Besides this, the radius corresponding to the luminosity and effective
temperature is larger than its measured (Roche-lobe) radius (cf.\ Kaper
2001). Thus, apart from being undermassive, the OB supergiants in HMXBs
also seem to be undersized for their luminosity and temperature. This
inconsistency is probably related to the phenomenon of Roche-lobe
overflow. The OB star tries to become a supergiant, but at some point it
reaches its critical Roche lobe and starts to transfer mass to its
companion. While the luminosity of the star is determined by the core
(which does not notice much of what is happening to the outer mantle),
the star would like to be bigger than allowed by its Roche lobe and is
peeled off.

\section{Neutron stars and black holes}

\begin{figure}
\centerline{\psfig{figure=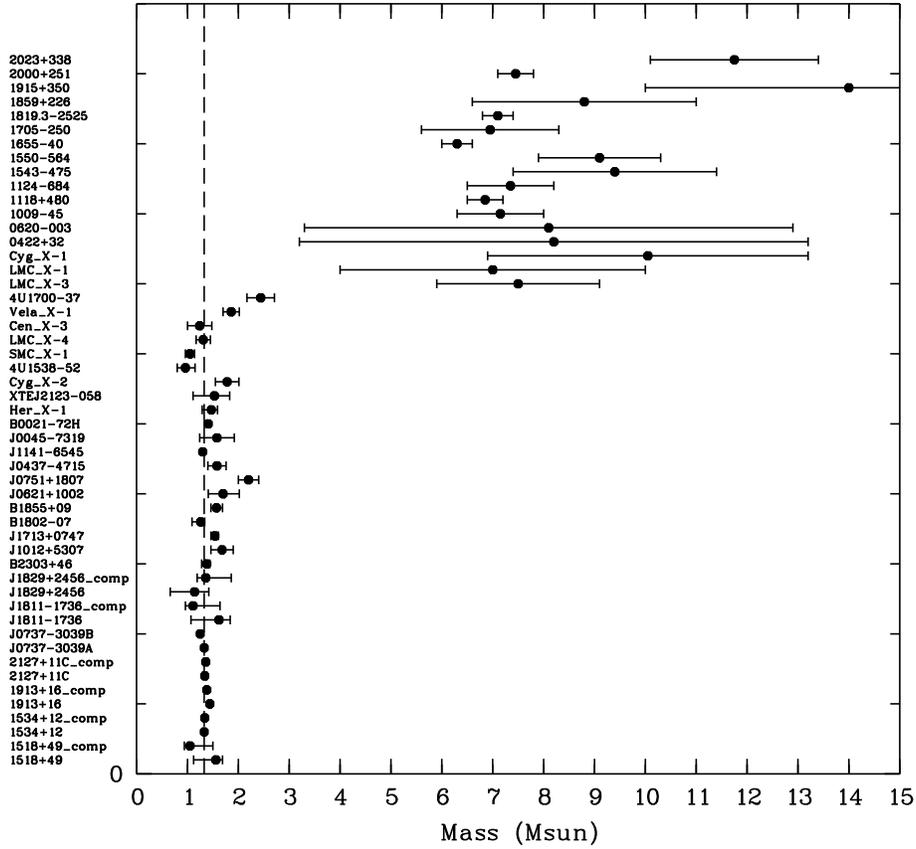,width=\textwidth,angle=-90}}
\caption[]{Neutron star and black hole masses obtained from literature
(Stairs 2004, McClintock \& Remillard 2005, and references
therein). The neutron stars, especially the binary radio pulsars (at
the bottom), occupy a relatively narrow mass range near
1.35~M$_{\odot}$. The X-ray pulsars (to the middle) show a wider
spread, including two systems with a neutron-star mass near
2~M$_{\odot}$. Such a high neutron star mass would rule out a soft
equation of state. The black-hole candidates (at the top) are
significantly more massive, indicative of a different formation
mechanism.}
\end{figure}

In Figure~1 the mass distribution is shown of neutron stars and black
holes, based on measurements collected from literature (Stairs 2004,
McClintock \& Remillard 2005). The neutron stars occupy a relatively
narrow mass range near 1.4~M$_{\odot}$. The most accurate neutron-star
masses have been derived for the binary radio pulsars, with an average
mass of 1.35$\pm$0.04~M$_{\odot}$ (Thorsett \& Chakrabarty 1999).  The
X-ray pulsars show a somewhat larger mass range, extending both below
and above 1.35~M$_{\odot}$. The neutron star in Vela~X-1 is
significantly more massive: $1.86 \pm 0.16$~M$_{\odot}$ (Barziv et
al. 2001, Quaintrell et al. 2003). Such a high neutron-star mass would
rule out the soft equations of state. Also for 4U~1700-37 a high
neutron-star mass is claimed ($2.4 \pm 0.3$~M$_{\odot}$, Clark et
al. 2002), although the X-ray source is, contrary to Vela~X-1, not an
X-ray pulsar (and perhaps a low-mass black hole).  Van der Meer et
al. (these proceedings) are currently analyzing the radial-velocity
curves of other OB-supergiant systems with an (eclipsing) X-ray pulsar
to find out whether Vela~X-1 is an exception or that the neutron stars
in these systems systematically deviate from the ``canonical'' mass of
1.35~M$_{\odot}$. This would provide an important constraint on the
neutron-star formation mechanism (i.e. the supernova).

Evolutionary calculations by Timmes et al. (1996) predict that massive
stars in close binaries which explode as Type~Ib supernova give rise
to initial neutron star masses in a narrow mass range around
1.3~M$_{\odot}$. This value does not include subsequent mass accretion
from a reverse shock or from a massive component in a binary system,
and Timmes et al. expect that the final masses could be somewhat
higher. Interestingly, for single stars, which explode as Type~II
supernovae, they find a bimodal neutron-star mass distribution, with
narrow peaks near 1.27 and 1.76~M$_{\odot}$. As mentioned, they did
not find a bimodal distribution for stars in close binaries, but at
present it is not clear whether this result will hold. If stars in
close binaries turn out to be more similar to single stars after all,
one could assign most neutron stars to the first peak, and Vela~X-1
(and 4U1700-37) to the second.

The estimated masses of black-hole candidates are substantially larger
($8.4 \pm 2.0$~M$_{\odot}$) than those measured for neutron
stars. This suggests that neutron stars and black holes are formed in
different ways. If, for example, black holes are the result of
``failed'' supernovae in which the stellar mantle is not blown away,
but accreting on the compact remnant, one would expect a significant
difference in mass between neutron stars and black holes. However, if
the mass of the (proto) neutron star is increased by the fall back of
material which was located outside the collapsing degenerate Fe core,
one would predict that neutron stars would occupy a range in mass, up
to the maximum neutron star mass allowed by the equation of
state. Certainly in the binary radio pulsars such a mass distribution
is not observed.  With the recent evidence that a black hole may be
formed during the collapse of a massive star during a gamma-ray burst
(GRB980425, Galama et al. 1998, Iwamoto et al. 1998), the hypothesis
would be that neutron stars are formed in ``ordinary'' supernovae,
while black holes originate from gamma-ray bursts.

\section{Microquasars}

Some X-ray binaries, most notably those hosting a black hole, produce
relativistic jets (e.g. Fender 2005). A famous example is SS433 which
shows collimated, precessing jets with velocities of $v = 0.26c$
(Margon 1982). In some systems, e.g. GRS1915+105, superluminal motions
have been measured, proving that the material in the jet is moving at
relativistic velocities as is observed in quasars. In our sample
(Tab.~1) three ``microquasars'' are included, i.e. LS5039, SS433, and
Cyg~X-1. LMC~X-1 and LMC~X-3 are also candidate members of this group,
but for these systems only upper limits are obtained in observations
searching for the radio synchrotron emission produced by the
jets. This jet phenomenon is not unique to black-hole systems; also
some X-ray binaries hosting a neutron star are known to produce jets
(e.g. Sco~X-1, Cir~X-1). Besides the jets, these systems also include
a (large) accretion disk. Apparently, a relatively large mass and
angular momentum accretion rate results in the formation of a dense
accretion disk and jets.

Cyg~X-1 is one of the most famous stellar-mass black-hole candidates
and one of the most intensively studied X-ray sources in the sky, at
all wavelengths. Cyg~X-1 probably represents a situation between pure,
spherical wind accretion and Roche-lobe overflow. Every few years
Cyg~X-1 makes a transition from a low/hard state to a high/soft state
in which the soft X-ray flux increases dramatically and the spectrum
softens for a period of weeks to months. The radio flux also varies
during state changes and is associated with jets (Stirling et
al. 2001).  The precise physical cause for the state transitions
remains unclear, but may be triggered by episodes of decreased
mass-loss rate in the supergiant donor star (Gies et al. 2003).

Hillwig et al. (2004) present spectroscopy of SS433 obtained near
primary eclipse and disk precessional phase $\Phi=0.0$, when the
accretion disk is expected to be most ``face-on''. These conditions
are the most favourable to have a change to detect the mass donor. The
spectra show clear evidence of absorption features consistent with a
classification of an A3-A7 supergiant. The observed radial velocity
variations are in antiphase to the disk spectrum; the latter includes
strong emission lines similar to those observed in Wolf-Rayet stars
(see also Fuchs et al., these proceedings). Hillwig et al. derive
masses of $10.9 \pm 3.1$~M$_{\odot}$ and $2.9 \pm 0.7$~M$_{\odot}$ for
the mass donor and compact object plus disk, respectively. 

LS5039 is an O6.5~V((f)) star (Clark et al. 2001) with a compact
companion, most likely a neutron star. It has radio-emitting
relativistic jets and is probably a high-energy gamma-ray source as
well (Paredes et al. 2000). It is a 4.4-day binary with a high
eccentricity ($e = 0.41$), which probably results from the huge mass
loss that occured with the supernova producing the compact
star. McSwain et al. (2004) present new optical and ultraviolet
spectra of the O star and find evidence for nitrogen enhancement and
carbon depletion in its atmosphere, indicative of the accretion of
nuclearly processed material originating from the compact star's
massive progenitor. The observed eccentricity and runaway velocity can
be reconciled only if the neutron star received a modest kick velocity
due to a slight asymmetry in the supernova explosion (during which
more than 5~M$_{\odot}$ was ejected).



\end{document}